# High Temperature Ferromagnetism with Giant Magnetic Moment in Transparent Co-doped $SnO_{2-\delta}$

S. B. Ogale[1,*], R. J. Choudhary[1], J. P. Buban[2], S. E. Lofland[3], S. R. Shinde[1], S. N. Kale[1], V. N. Kulkarni[1,#], J. Higgins[1], C. Lanci[3], J. R. Simpson[4], N. D. Browning[2], S. Das Sarma[4], H. D. Drew[4], R. L. Greene[1] and T. Venkatesan[1].

[1]Center for Superconductivity Research, Department of Physics, University of Maryland, College Park, MD 20742-4111.

[2]Department of Physics, University of Illinois at Chicago, 845 West Taylor Street, Chicago, IL 60607-7059.

[3]Department of Chemistry and Physics, Rowan University, Glassboro, N.J. 08028-1701.

[4]Department of Physics, University of Maryland, College Park, MD 20742-4111.

Occurrence of room temperature ferromagnetism is demonstrated in pulsed laser deposited thin films of $Sn_{1-x}Co_xO_{2-\delta}$ (x<0.3). Interestingly, films of $Sn_{0.95}Co_{0.05}O_{2-\delta}$ grown on R-plane sapphire not only exhibit ferromagnetism with a Curie temperature close to 650 K, but also a giant magnetic moment of $7 \pm 0.5$ $\mu_B$/Co, not yet reported in any diluted magnetic semiconductor system. The films are semiconducting and optically highly transparent.

Integrating spin functionality into otherwise non-magnetic condensed matter systems has become a highly desirable goal in the context of the rapidly developing field of spintronics [*1-5*]. The ability to modulate the electrical and optical properties of systems by spin control not only lends itself to the development of innovative applications, but is also of significant interest from the fundamental standpoint of structure-chemistry-property relationships in solids.

One of the apparently straightforward means of introducing spin effects in non-magnetic solids is to introduce magnetically active dopants such as transition elements into the matrices and hope that they not only remain magnetically active but also couple with the electronic states of the solid. Considerable success has been achieved in inducing ferromagnetism by transition element doping in compound semiconductor systems [*1-4*] although the Curie temperatures are much lower than room temperature. In the case of the oxide systems, the efforts as well as successes in this context are still relatively limited [*6-11*] in spite of the fact that many strongly correlated oxides provide interesting grounds for a complex interplay of charge, spin and orbitals. Early attempts at the synthesis of diluted magnetic semiconductor (DMS) oxides used ZnO as the host and yielded mixed results [*6-8*]. A rather strong indication of favorable results was subsequently obtained in the substrate-stabilized anatase film phase of Co doped $TiO_2$ system [*9,10*]. Questions regarding the precise state of cobalt and local microstate of the host are however still being sorted out [*11*].

In this work we use $SnO_2$ as the host matrix in view of its already well documented interesting optical and electrical properties. This material has widespread applicability [*12-15*] in fields such as gas sensing (even for flammable and toxic gases), transparent conducting electrodes in flat-panel displays and solar cells, IR detectors, optoelectronic devices etc. It also has superior chemical stability in comparison with other wide gap semiconductors [*16*]. While most DMS systems studied so far are hole doped, $SnO_2$ has an n-type conduction [*17*]. For spintronic device application electron-doped magnetic semiconductors are essential. This system, which configures in rutile phase, has been recently explored with Mn doping but no ferromagnetism has been reported [*18*]. By doping this material with cobalt we show that this system in its pulsed laser deposited thin film form leads to high temperature ferromagnetism with a Curie temperature as high as 650 K, while still retaining its highly desirable optical transparency and semiconductivity. Remarkably, at low dopant concentration, a giant magnetic moment of $7 \pm 0.5$ $\mu_B$/Co is also observed, which has not been seen in any DMS system so far.

The ceramic targets used for pulsed laser ablation were prepared by standard solid state reaction technique. The depositions were performed at a substrate temperature of 700 °C and oxygen partial pressure of $1 \times 10^{-4}$ Torr. The laser energy density and pulse repetition rate were kept at 1.8 J/cm$^2$ and 10 Hz, respectively. The samples were cooled in the same pressure as used during the deposition, at the rate of 20 °C/min. It was found that the cobalt content in the film is considerably higher than that in the corresponding target, possibly due to partial evaporation of low melting point material Sn from the growth front. The compositions quoted for the films are therefore the actual concentrations in the film obtained by Rutherford backscattering spectroscopy (RBS). The films were characterized by x-ray diffraction (XRD), scanning transmission electron microscopy (STEM), SQUID and VSM magnetometries, transport and optical measurements, and ion channeling.

Fig. 1 (a) shows XRD pattern for a $Sn_{0.95}Co_{0.05}O_{2-\delta}$ thin film grown on R-plane sapphire substrate. Only the (101) family of rutile phase film peaks is seen, as expected. The rocking curve (inset) full width at half maximum of $0.23^o$ signifies excellent orientational quality. Interestingly, similar good XRD signatures were obtained even for films with cobalt concentrations as high as 30% (x=0.3). A decrease in the XRD peak intensity was encountered only for further increase in the cobalt content, as shown in Fig. 1 (b). Shown in the inset of Fig. 1(b) is a high resolution STEM image for a $Sn_{0.73}Co_{0.27}O_{2-\delta}$ film, indicating that even up to such high cobalt concentration the film microstructure is uniform. Indeed, in Fig. 1 (c) we show the EELS data recorded at various points spread over the TEM image domain in order to establish the chemical uniformity. One can clearly see that cobalt is distributed uniformly in the film.

RBS ion channeling data recorded for cobalt lattice location (not shown) exhibited a fairly good channeling for Sn, but hardly any channeling for Co. Since the TEM data do not show any indication of clustering of the dopant, the lack of channeling of cobalt implies either its interstitial nature or symmetry along a different axis or substitutionality with local distortions in the site leading to an incoherent distribution [*19*]. The incoherency could arise due to the oxygen vacancies in the proximity of the dopant presumably formed to account for its lower valence as compared to Sn. Since the vacancies may form in any of the neighboring oxygen sites, different cobalt atoms could be displaced randomly along different directions.

In Fig. 2 (a) we show magnetic hysterisis loops at room temperature for $Sn_{0.95}Co_{0.05}O_{2-\delta}$ and $Sn_{0.73}Co_{0.27}O_{2-\delta}$ films. A well defined hysterisis loop with (coercivity ~50 Oe) is seen in each case. Remarkably, for the case of the $Sn_{0.95}Co_{0.05}O_{2-\delta}$ film a giant magnetic moment as high as $7 \pm 0.5$ $\mu_B$/Co is observed, and it is seen to drop rapidly with increase in the cobalt content, as shown in the inset. To our knowledge no such giant moment has ever been reported in any DMS system. Giant moments have been observed earlier in transition metal doped palladium and alkali metal systems [*20-26*], and have been a subject of interesting scientific analyses and debate for many years. Realization of such large moment in an optical material could be useful from the applications standpoint. We will return to the possible origin of such a giant moment and its concentration dependence.

In Fig. 2 (b) we show the magnetization as a function of temperature for the $Sn_{0.95}Co_{0.05}O_{2-\delta}$ film measured from 4.2 K to 300 K using SQUID and from 300 K to 700 K using VSM. The magnetization is seen to be fairly constant up to about 475 K, and begins to drop thereafter leading to a Curie temperature close to 650 K. Interestingly, after the sample was cooled to room temperature, the magnetic moment was seen to have dropped by a factor of about 3. It can thus be inferred that the state of the as-grown film with a giant moment is a metastable one. In the inset of Fig. 2(b) we show the magnetization data for the $Sn_{0.92}Co_{0.08}O_{2-\delta}$ sample in the high temperature region. As compared to the case of the x=0.05 Co-doped sample, this x=0.08 Co-doped sample has a broader tail with non-zero moment extending above 650 K, the possible origin of which will be discussed later.

In Fig. 3 we compare the optical conductivity for the undoped $SnO_{2-\delta}$ and $Sn_{0.73}Co_{0.27}O_{2-\delta}$ films. The transparency in the gap region ($\omega < 3.8$ eV) is seen to be virtually unaffected even up to such high cobalt concentrations. This is illustrated in the inset to Fig. 3. In addition there is no evidence for impurity levels in the gap as might be expected from the partially filled d-states in cobalt. The absorption edge is seen to shift to higher wavelength (lower energy) for cobalt doped sample, indicating matrix incorporation of cobalt atoms. Such shifts are known to occur in DMS systems and depend on the particular host and dopant concentration [*27-28*]. Further work to explore these aspects experimentally and theoretically is now in progress.

In Fig. 4 (a) we compare the transport properties of the undoped and Co-doped $SnO_{2-\delta}$ films. While the room temperature resistivity for the undoped film under specified growth conditions is 0.03 Ω-cm, it jumps to about 0.4 Ω-cm upon x=0.05 cobalt doping. In films with x=0.15 and x=0.27 cobalt, the room temperature resistivity is about 200 and 4000 Ω-cm, respectively. The $Sn_{0.95}Co_{0.05}O_{2-\delta}$ film shows a rapid increase in resistivity below about 10 K, possibly due to trapping of the carriers into shallow impurity related traps. This low temperature resistivity behavior also has interesting manifestations in the magnetoresistance (MR = $(\rho_H-\rho_0)/\rho_0$ x 100%, $\rho_0$ and $\rho_H$ being the resistivity without and with field H).

In Fig. 4 (b) we show the MR as a function of field up to 14 tesla for the $Sn_{0.95}Co_{0.05}O_{2-\delta}$ film at 4, 10 and 20 K. One can see that the MR is strong and positive at 4 K, while relatively weak and negative at 20 K. At 10 K it shows a non-monotonic field dependence. The positive MR at very low temperature which corresponds to carrier trapping and excitation regime from the shallow traps implies Zeeman splitting of these states due to their strong coupling to the moment of cobalt causing the deepening of the traps. Such shallow states may originate in oxygen vacancies and their coupling to cobalt moment may occur because of the need of the cobalt atom to have a nearby oxygen vacancy for satisfying its lower valence compared to that of Sn which is $4^+$.

It is now useful to make a few remarks on the giant ($7 \pm 0.5$ $\mu_B$/Co) magnetic moment observed in the $Sn_{0.95}Co_{0.05}O_{2-\delta}$ film in the as-grown state. This value of the moment is much larger than the value of about 1.67 $\mu_B$/Co for the case of cobalt metal, or that for small cobalt clusters for which a moment of about 2.1 $\mu_B$/Co is possible [*29,30*], or that of any of the standard cobalt oxides wherein the orbital moment is quenched. One possibility is that the atoms surrounding the cobalt atoms have acquired a moment through electronic effects, or the orbital moment of cobalt remains unquenched. Magnetic moments significantly larger (about 6-16 $\mu_B$/atom) than the spin-only-moments have indeed been reported in transition metal atoms doped in or spread on the surfaces of alkali metal solids such as Cs [*20-26*], and these have been attributed to the unquenched orbital contributions. In such cases, increase in the concentration of dopants has been found to cause a rapid decrease in the moment (as observed in our sample) due to enhanced dopant-dopant associations leading to progressive orbital moment quenching. The decreased moment observed in our higher Co-doped samples, and the drop in the moment after a high temperature treatment in low Co-doped sample, possibly caused by enhanced associations, suggest that the scenario may be similar in our case. The tail extending beyond 650 K in the temperature dependence of magnetization for the x=0.08 Co-doped sample also suggests increased dopant association. A clear understanding of these issues will require further work.

In conclusion, thin films of $Sn_{0.95}Co_{0.05}O_{2-\delta}$ grown by pulsed laser deposition on single crystal sapphire substrates are seen to be ferromagnetic, with a Curie temperature close to 650 K and a giant magnetic moment of $7 \pm 0.5$ $\mu_B$/Co. Such a giant moment suggests that either the cobalt orbital moment is not quenched or some moment appears on the neighbors of cobalt in the matrix. The films are highly transparent and semiconducting. Films with higher cobalt content (up to ~ x=0.30) are also ferromagnetic and transparent, but are highly resistive and have a much lower magnetic moment. No cobalt clustering is observed in high resolution scanning transmission electron microscopy, but cobalt is found to be occupying a sublattice with a different axis of symmetry or possibly in an incoherent lattice configuration.

This work is supported by DARPA SpinS program and by NSF-DMR-MRSEC and NSF-ECS-EPDT. The authors like to thank Chris Lobb and Steven Anlage for critical reading of the manuscript.

Figure Captions :

Fig. 1 (a) X-ray diffraction (XRD) pattern for a $Sn_{0.95}Co_{0.05}O_{2-\delta}$ thin film grown on R-plane sapphire substrate. The inset shows the rocking curve; (b) Normalized (101) XRD peak intensity as a function of the cobalt content. The inset shows a high resolution STEM image for a $Sn_{0.73}Co_{0.27}O_{2-\delta}$ film; (c) Electron Energy Loss Spectra (EELS) recorded at various points spread over the TEM image domain of the image shown in (b). The spectra show the Sn-M, O-K and Co-L edges. They are shifted on the y-scale for clarity.

Fig. 2 (a) Magnetic hysterisis loops for $Sn_{0.95}Co_{0.05}O_{2-\delta}$ and $Sn_{0.73}Co_{0.27}O_{2-\delta}$ films at 300 K. Inset shows the dependence of saturation moment $M_S$ on cobalt concentration; (b) Magnetization as a function of temperature for the $Sn_{0.95}Co_{0.05}O_{2-\delta}$ film measured from 4.2 K to 300 K using SQUID and from 300 K to 700 K using VSM. Inset shows the VSM data for the $Sn_{0.92}Co_{0.08}O_{2-\delta}$ film. The dashed lines in the inset are guide to the eye.

Fig. 3 Optical conductivity ($\sigma_1$) in undoped $SnO_{2-\delta}$ and $Sn_{0.73}Co_{0.27}O_{2-\delta}$ films. The inset shows the transparency of a $Sn_{0.73}Co_{0.27}O_{2-\delta}$ film in the visible range.

Fig. 4 (a) Temperature dependence of resistivity for the $Sn_{1-x}Co_xO_{2-\delta}$ films; (b) The magnetoresistance (MR) as a function of magnetic field upto 14 tesla for the $Sn_{0.95}Co_{0.05}O_{2-\delta}$ film at 4, 10 and 20 K.

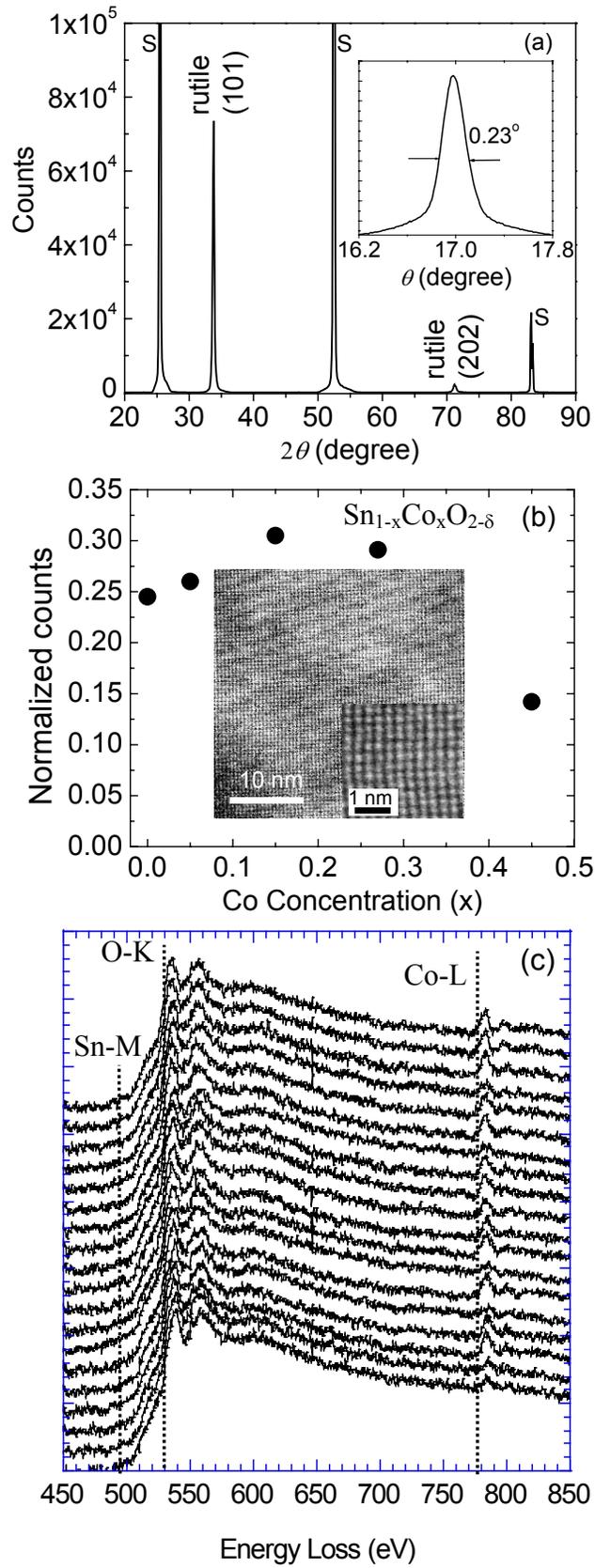

Figure 1: Ogale et al.

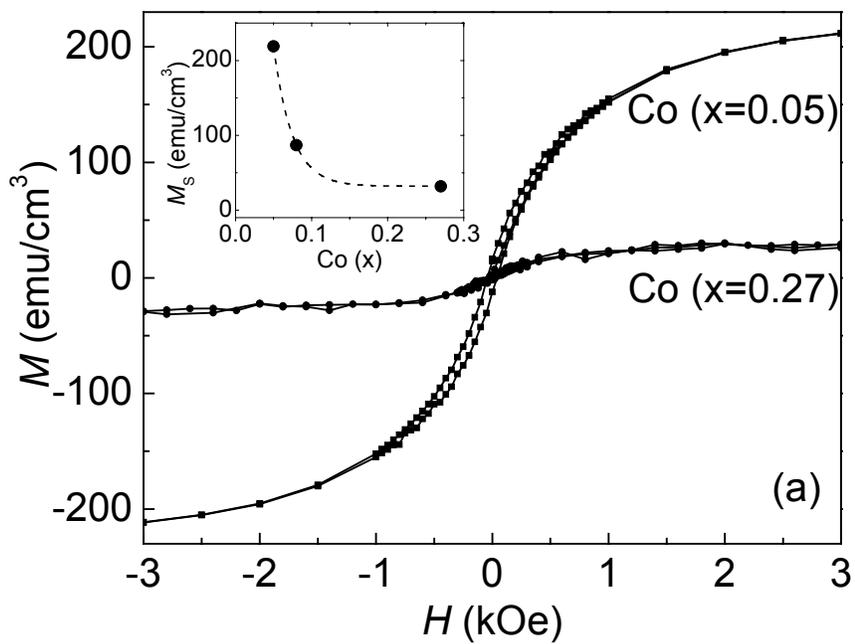

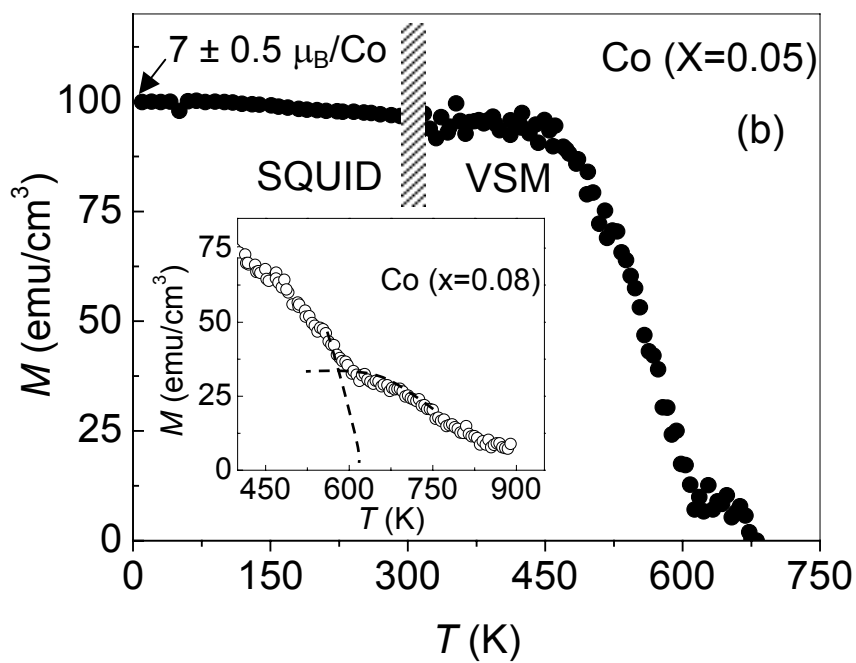

Figure 2: Ogale et al.

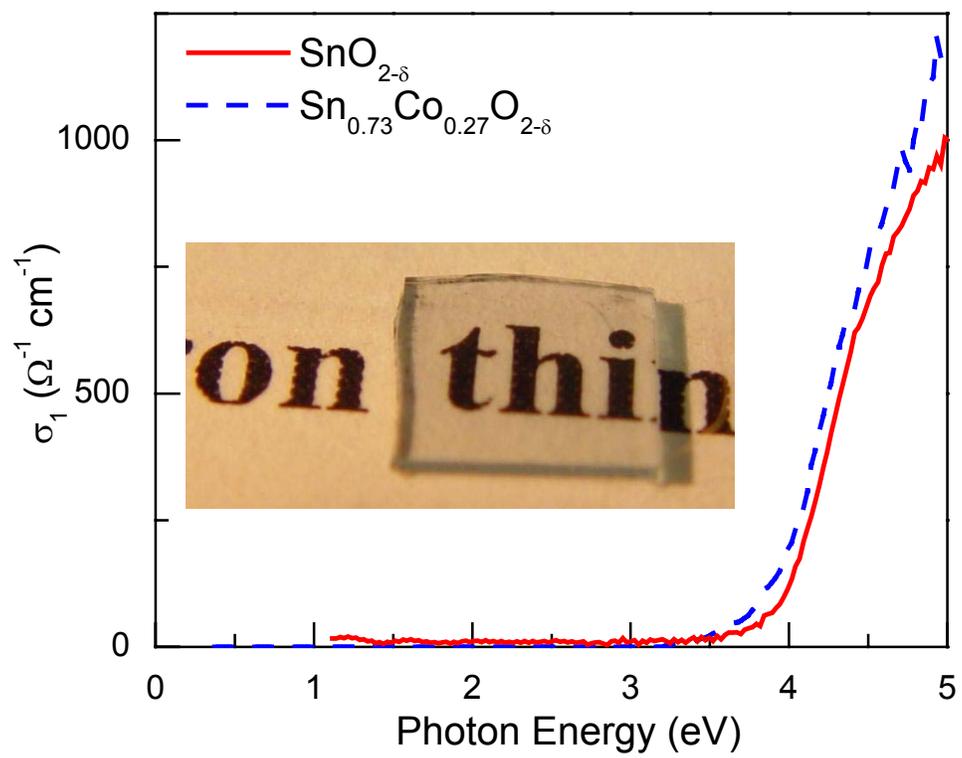

Figure 3: Ogale et al.          15

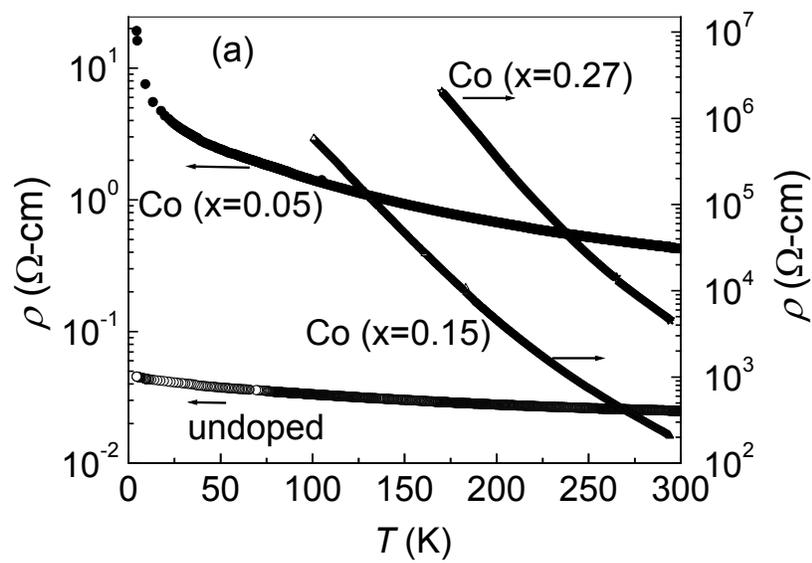
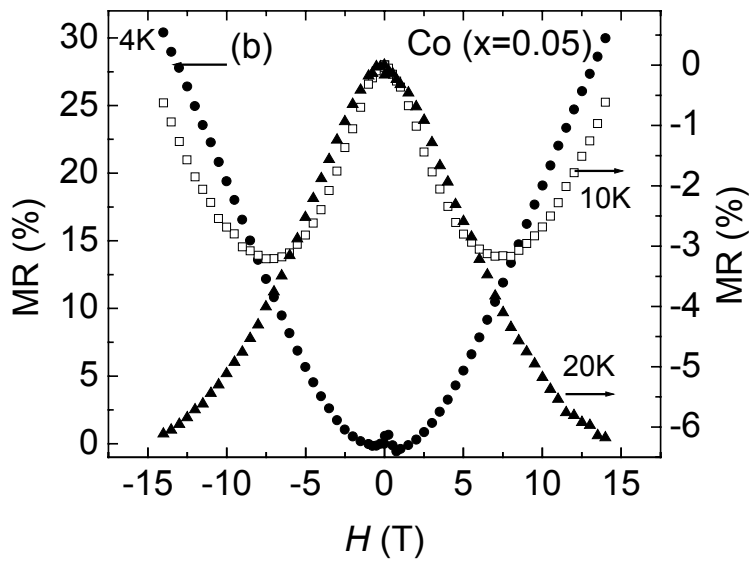

Figure 4: Ogale et al.                    16